\documentclass[twocolumn,eqsecnum,showpacs,showkeys,amsmath,amssymb,nofootinbib,superscriptaddress,floatfix]{revtex4}

\usepackage{graphicx}
\usepackage{subfigure}
\usepackage{bm}

\makeatletter
\newcommand\erfc{\mathop{\operator@font erfc}\nolimits}
\def\slashchar#1{\setbox0=\hbox{$#1$}
   \dimen0=\wd0 \setbox1=\hbox{/} \dimen1=\wd1
   \ifdim\dimen0>\dimen1 \rlap{\hbox to \dimen0{\hfil/\hfil}} #1
   \else  \rlap{\hbox to \dimen1{\hfil$#1$\hfil}} / \fi}

\makeatother

\begin{document}
\title{Interplay of the emission from
 thermal and direct sources in relativistic heavy 
ion collisions\footnote{Supported by 
Polish Ministry of Science and Higher Education under
grant N202~034~32/0918}}
\author{Piotr Bo\.zek}
\email{Piotr.Bozek@ifj.edu.pl}
\affiliation{Institute of Physics, Rzesz\'ow University, PL-35959 Rzesz\'ow, Poland}
\affiliation{The H. Niewodnicza\'nski Institute of Nuclear Physics,
PL-31342 Krak\'ow, Poland}
%

\begin{abstract}
The separation of the source created in ultrarelativistic 
heavy-ion collisions into a thermalized dense core and 
an outer mantle consisting of independent nucleon-nucleon 
collisions is discussed. Evidence for such a two component picture is found in 
transverse mass spectra of kaon, protons and antiprotons 
produced  in Au-Au collisions at $\sqrt{s}=200$GeV.  Estimates of the sizes 
of the thermal and direct sources are compared to models separating the
 interaction zone into a core and a corona, according to  the density of
 participants or to the number of collisions. Consequences for the modeling 
of the dynamics of the small size,  thermalized core are described. 
 New initial conditions corresponding to the dense core 
 lead to a stronger azimuthal asymmetry of the 
hydrodynamically expanding fireball, pressure gradients  also increase.  
$2\mbox{+}1$-dimensional hydrodynamic
simulations are presented starting from all the matter in the 
interaction region or from the dense, thermal part of the source. 
We find faster transverse expansion and stronger elliptic flow for 
dense core initial conditions. For different impact parameters 
we find very similar spectra of the thermal part of the 
source and only adding  particles emitted directly from nucleon-nucleon
collisions in the corona the experimentally observed 
softening of the spectra  with increasing impact parameter is reproduced. The 
elliptic flow is stronger for particles emitted 
 from a source separated into a core and a corona.
\end{abstract}



\maketitle

\section{Introduction}

A vast number of experimental observations indicate that 
matter created in relativistic heavy ion collisions is a strongly 
interacting, dense and thermalized medium 
\cite{Adams:2005dq,Adcox:2004mh,Arsene:2004fa,Back:2004je}. 
Transverse momentum spectra of emitted particles are thermal. 
 Fitted  temperatures are in the range of $100-160$MeV and the transverse
 velocity of the emitting source is $0.5-0.6$c on average. The presence
 of  a substantial transverse flow confirms the formation of a dense
 matter that expands collectively.  Since no transverse flow is present 
in the initial stage of the collision, the observed expansion results from a
 build up of collective flow from density gradients in the fireball. The 
expansion continues until particles decouple from the system, i.e. until
 freeze-out. Another important achievement is the experimental observation 
of azimuthal asymmetry in the collective flow.
Spectra of particles emitted at central rapidities are written
 using the elliptic flow coefficient $v_2$,
\begin{equation}
\frac{dN}{p_\perp dp_\perp d\phi}=\frac{dN}{2\pi p_\perp dp_\perp}
\left(1+v_2(p_\perp) \cos (2\phi)\right) .
\end{equation}
The parameter $v_2$ has been measured for a variety of identified 
particles and  transverse momenta. For non-central collisions substantial
elliptic flow is observed, increasing with transverse momenta up 
to $p_\perp \simeq 1.5$GeV. The azimuthal asymmetry in this kinematical
 range can be explained by the  collective expansion of an azimuthally 
asymmetric source followed by thermal emission at freeze-out.
Hanbury Brown-Twiss  (HBT) correlations between identical
 particles allow to estimate the lifetime and the size of the fireball. 
Correlation analysis confirms the existence of a strong collective flow and 
suggest a fast  expansion of the system.

Thermal fits \cite{Schnedermann:1993ws} of spectra of 
emitted particles show an increase of the radial
 flow with centrality and a decrease of the temperature \cite{Adams:2003xp}.
The effect manifest itself as well as an increase of the average $p_\perp$ 
of emitted particles with centrality. 
 Observed ratios of the number 
of produced particles can be calculated  assuming 
 particle production in a state of 
chemical equilibrium, 
defined  by the  temperature and the values of
 chemical potentials 
\cite{Andronic:2005yp,Rafelski:2004dp,Cleymans:2004pp,Becattini:2005xt,Florkowski:2001fp}.
For central collisions at $\sqrt{s}=200$GeV particle 
ratios follow chemical equilibrium,  but
 when going to more peripheral collisions the relative number of strange 
particles produced decreases. This decrease is taken into account
 by the  introduction of a strangeness suppression factor $\lambda_s \le 1$ 
that reaches $1$  for central collisions.

Properties of  the hot, dense and strongly interacting fireball can be 
modelled  by relativistic fluid dynamics (for reviews see 
\cite{Kolb:2003dz,Nonaka:2007nn,Huovinen:2006jp,Hirano:2008hy}). Matter
is usually  assumed to be a perfect fluid, although possible effects 
of viscosity and other sources of dissipation are discussed.
Hydrodynamic models qualitatively describe  the dynamics 
of the hot source created in the collision. Fluid dynamic models include  
several parameters and assumptions~: initial starting time and 
initial density profile, the equation of state (EOS), the freeze-out conditions 
and sometimes shear viscosity. Only recently a consistent description of soft
 observables measured in heavy ion collisions at RHIC energies has been 
achieved \cite{Chojnacki:2007rq,Broniowski:2008vp,Kisiel:2008ws}. 
Assuming an early thermalization and an EOS without a soft-point, transverse 
momentum spectra, elliptic flow,  HBT  radii, and azimuthally
 sensitive HBT radii could be described in central and
 semi-peripheral collisions. The dependence of the elliptic flow on the impact 
parameter and in particular its scaling properties \cite{Voloshin:2007af} 
indicate that an almost perfect fluid is formed with 
possible deviations due to dissipative effects in most peripheral collisions.

The dependence of  strangeness production, 
average $p_\perp$, spectra, transverse and elliptic flows on centrality
 is due to a change of the size and shape of the fireball. 
To explain the observed systematics one has to invoke different 
freeze-out conditions for collision events in different centrality classes. 
While some reduction of the transverse flow with the decreasing
 size of the system is possible, a significant change in the 
freeze-out temperature is not understood. However,
recently the effect of energy and momentum conservation
 on the fitted temperatures has been noticed \cite{Chajecki:2008yi}.
 The reduced strangeness production in 
peripheral collisions calls for the introduction of a new parameter,
 the strangeness suppression factor, describing the degree to which chemical 
equilibration is achieved. On the other hand, a simple idea assuming that
 the interaction zone in heavy-ion collisions is composed of two 
sources explains the observed impact-parameter dependence \cite{Bozek:2005eu}.
The most dense part of the interaction zone is the thermal fireball,
 that behaves collectively and is chemically and thermally equilibrated.
The outer mantle of the interaction zone does not take part in the fireball 
formation. Particles emitted from this part of the system do not reinteract 
significantly
and their spectra and abundances are the same as in  nucleon-nucleon (N-N) 
collisions.  Assuming a minimal density of  participant nucleons in the 
transverse plane necessary for the local thermalization  \cite{Bozek:2005eu}, 
we could reproduce the centrality dependence of the multiplicity 
of charged particles and of the strange particle suppression ($K/\pi$ ratio) 
on centrality. The thermal source emits more particle per participant
 nucleon pair than a single N-N
 collision.  In the most central collisions 95\% 
of the participant nucleons end up in the thermal fireball and only 5\% 
of them are in the corona, on the other hand for centralities
 70-80\% the corona dominates the emission. The change in the proportion of 
the thermal fireball and of the corona with centrality leads to 
a stronger than linear increase of the particle multiplicity with
 the number of participants. A very similar mechanism can explain the 
change in the ratio of strange particles to other particles with the centrality
of the collision. We assume that the thermal fireball is always close to a 
complete chemical equilibrium, but its proportion in the total particle 
emission goes down in peripheral collisions, as relatively more particles 
are emitted
from the outer mantle. The rate of the emission of  
 strange particles  from the corona is below chemical equilibrium.
The assumption \cite{Pantuev:2005jt} 
 that jet absorption is stronger in the dense thermal 
fireball leads to results consistent with the experimental 
data on nuclear attenuation rate  \cite{Bozek:2005eu}. It has been noticed that
 the spectra of particles emitted from the two components in the interaction 
region have different slopes \cite{Werner:2007bf}. For central collisions,
with increasing
 proportion of particles emitted from the thermal fireball, the 
spectra become harder. Recently, Becattini and  Manninen 
\cite{Becattini:2008yn} analyzed strangeness suppression 
in peripheral nuclear collisions for different particle species
 assuming a two component source. They used a different definition of 
the thermalized fireball based on the number of collisions that a participant
 undergoes.

In the present work we analyze the two-component picture of the interaction
 zone and its consequences on the spectra and elliptic flow of emitted 
particles. In Sect. \ref{sec:fit} we perform a fit of the
 thermal core and corona sizes for different centralities in order to reproduce
the observed spectra of kaons, protons and antiprotons. We compare 
the extracted thermal fireball and corona sizes for different centralities
 with models dividing the interaction region according to the density
of participants in the transverse plane  \cite{Bozek:2005eu} 
or according to the number of collisions \cite{Becattini:2008yn}.
In Sect. \ref{sec:dense} we look at  the geometry
of the thermal fireball for different impact parameters, and compare it to 
the geometry of the fireball calculated  in the usual Glauber model.
Then in Sect. \ref{sec:hydro} a hydrodynamic evolution of the thermal fireball
is performed ending with a freeze-out emission of particles. 
Calculated spectra and elliptic flow  are compared to results of a
hydrodynamic expansion 
starting from a fireball consisting of the whole interaction region.
 We find that spectra of particles  emitted from an expanding thermal, dense 
fireball are harder and show a stronger elliptic flow, than standard
 hydrodynamic model predictions. The differences get partially 
reduced when emission from the corona is added to the final spectra.

\section{Two components in the spectra}
\label{sec:fit}

\begin{figure}
\includegraphics[width=.4\textwidth]{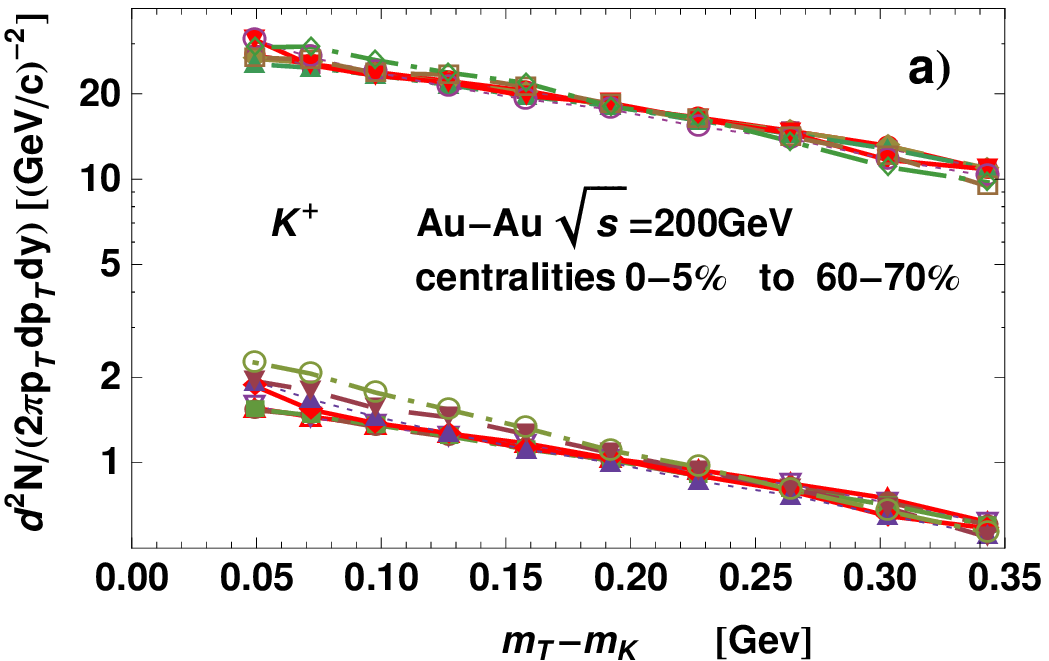}
\includegraphics[width=.4\textwidth]{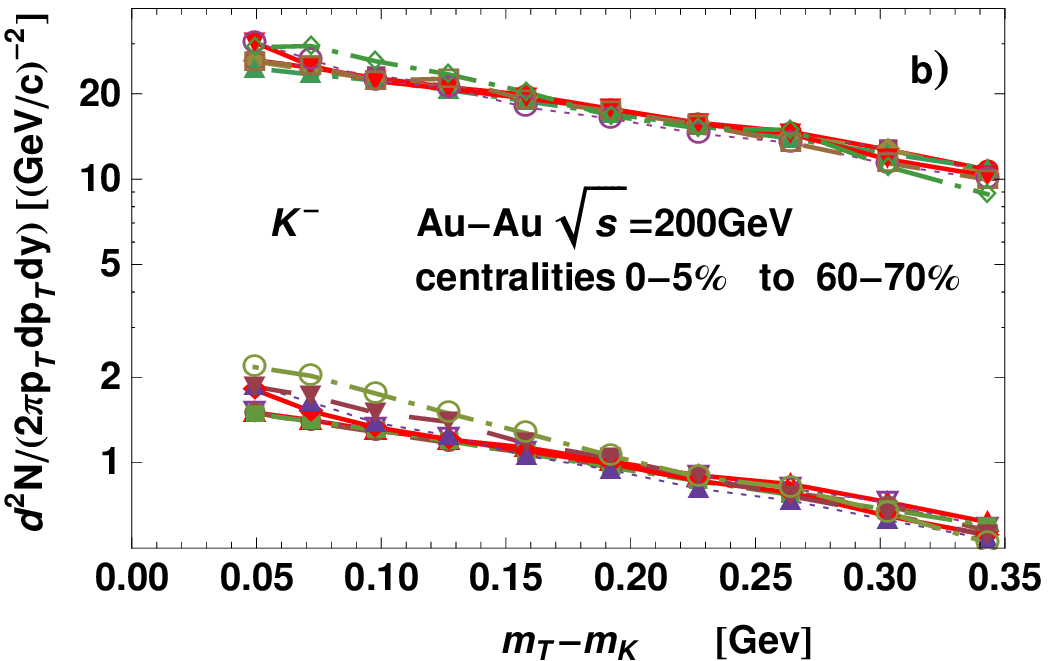}
\caption{(Color online) The upper points represent the 
thermal part of the spectra for $K^+$ (panel a)) and $K^-$ (panel b))
 for Au-Au collisions at $\sqrt{s}=200$GeV 
(STAR Collaboration data \cite{Adams:2003xp}). Spectra are obtained by the
fit procedure consisting in 
 subtracting the contribution from the corona 
and by  scaling to the size of the source in $0\mbox{-}5$\% centrality bin.
 Results 
for 8
  centrality bins from $0\mbox{-}5$\% to $60\mbox{-}70$\% are superimposed.
 The lower curves represent the raw spectra in different centrality classes 
divided by the number of participant 
pairs $N_{part}/2$ ($\times 10$).}
\label{fig:kaons}
\end{figure}

Experimental data on identified particle spectra at different impact parameters 
show a hardening of the spectra in transverse momentum when going 
to central collisions. The effect is more pronounced for heavy particles, 
which indicates the role of the  transverse flow therein.
  In the two-component model of the emission, 
particles originate from two different zones, the thermal fireball 
and the outer corona. The dominant reason for the change of the
spectra with centrality is not the change of the parameters of the thermal 
source (freeze-out temperature and velocity) but the change in the 
proportion of  particles emitted from the thermal source and from
 independent N-N collisions in the corona.

\begin{figure}
\includegraphics[width=.4\textwidth]{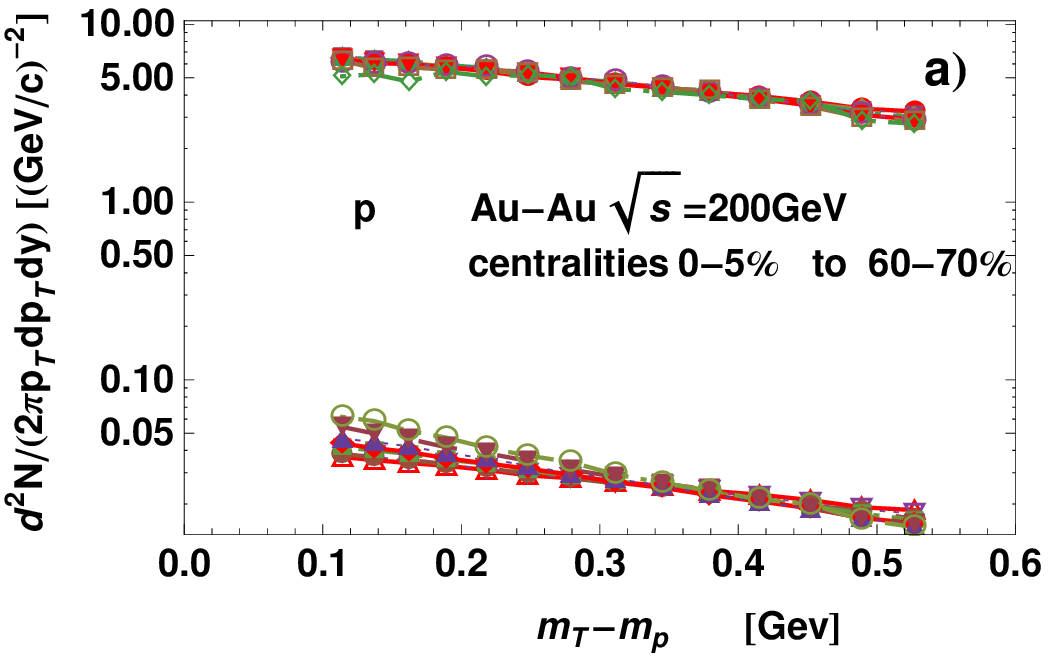}
\includegraphics[width=.4\textwidth]{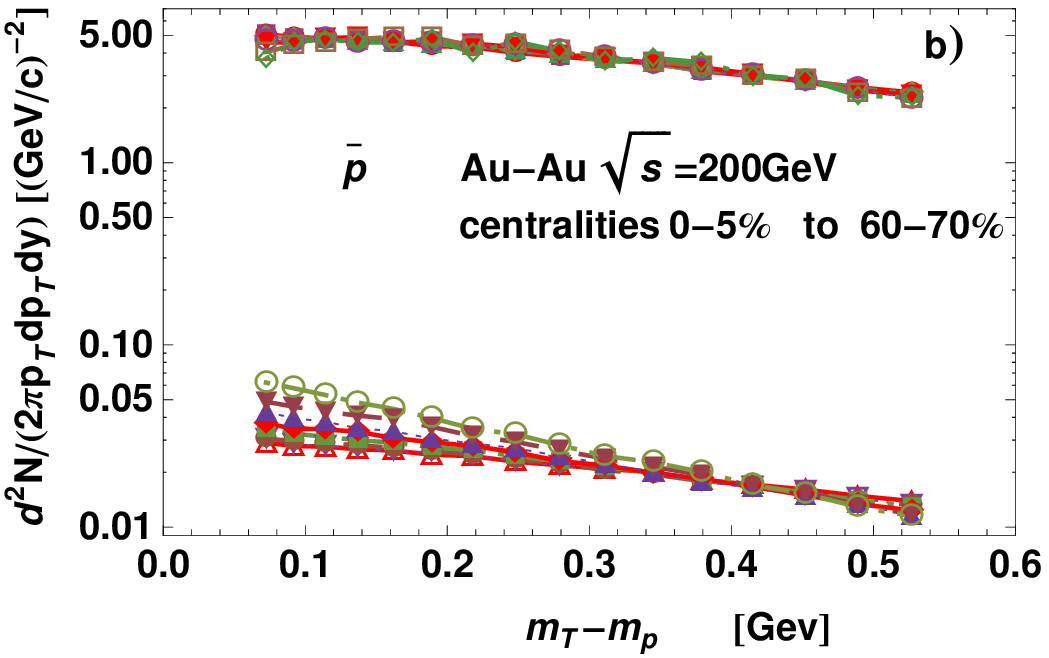}
\caption{(Color online) Upper points represent the extracted
 thermal part of the spectra for protons (panel a)) and antiprotons 
(panel b)) similar as for kaons in Fig. \ref{fig:kaons}. Lower points
 represent the raw measured spectra in different centralities divided  by
 the number of participant 
pairs $N_{part}/2$.  }
\label{fig:pap}
\end{figure}

For each particle type we have
\begin{eqnarray}
\label{eq:twospectra}
\frac{dN(c)}{2\pi p_\perp d p_\perp dy}&=&\frac{N_{core}(c)}{2}\frac{dN_{th}}
{2\pi p_\perp d p_\perp dy}\nonumber \\
&+&\frac{N_{corona}(c)}{2}
\frac{dN_{pp}}{2\pi p_\perp d p_\perp dy} 
\end{eqnarray}
where $N_{core}(c)$ and $N_{corona}(c)$ are the number of participants in the 
core and in the corona at a given centrality $c$,
 $dN_{pp}/2\pi p_\perp d p_\perp dy$ is the
 spectrum in proton-proton collisions and  
$dN_{th}/2\pi p_\perp d p_\perp dy$ is the spectrum of particles emitted from 
the thermal fireball per participant pair. The proton-proton spectra 
$dN_{pp}/2\pi p_\perp d p_\perp dy$ 
have been measured experimentally \cite{Adams:2003xp}
and can be subtracted from the 
raw spectra on the left hand side of Eq. (\ref{eq:twospectra}). The thermal 
spectra $dN_{core}/2\pi p_\perp d p_\perp dy$ 
correspond to particles emitted from the thermal fireball after
 expansion
 and freeze-out, we assume that they depend weakly on the centrality.
Spectra written below after  subtraction  and  rescaling
  should not depend on centrality
\begin{eqnarray}
\label{eq:prefit}
& & \frac{N_{core}({0\mbox{-}5\%})}{N_{core}(c)} \nonumber \\
& & \left[\frac{dN(c)}{2\pi p_\perp d p_\perp dy}-
\frac{N_{corona}(c)}{2}
\frac{dN_{pp}}{2\pi p_\perp d p_\perp dy} \right]  \ ;
\end{eqnarray}
they  are equal to the spectra of particles from the thermal
 fireball in the most central bin ($0\mbox{-}5$\%).
For each centrality other than the most central bin Eq. (\ref{eq:prefit}) 
represents a constraint on the parameters $N_{core}$ and $N_{corona}$. These
 parameters should be adjusted  to make the spectrum in (\ref{eq:prefit})
 as close as possible to 
\begin{eqnarray}
\label{eq:fit2}
\frac{N_{core}({0\mbox{-}5\%})}{2}\frac{dN_{th}}{2\pi p_\perp d p_\perp dy} =
& &\nonumber \\
\frac{dN(0\mbox{-}5\%)}{2\pi p_\perp d p_\perp dy}-
\frac{N_{corona}({0\mbox{-}5\%})}{2}
\frac{dN_{pp}}{2\pi p_\perp d p_\perp dy} & & \ .
\end{eqnarray}
The fit cannot separate the core and 
corona contributions for one of the centralities, in our case $0\mbox{-}5$\%.
Using the data of the STAR Collaboration \cite{Adams:2003xp,Abalev:2008ez}, 
for 7 centralities $5\mbox{-}10$\%, $10\mbox{-}20$\%, \dots to 
$60\mbox{-}70$\%, we perform a $\chi^2$
 fit for two parameters
$\alpha(c)$ and $\beta(c)$ so that the difference  between the two
 sides of the
following equation
\begin{eqnarray}
\label{eq:albe}
& &N_{core}({0\mbox{-}5\%}) \frac{dN_{th}}{2\pi p_\perp d p_\perp dy} \nonumber \\
& & \simeq \left(\alpha(c)\frac{dN(c)}
{2\pi p_\perp d p_\perp dy} -\beta(c)\frac{dN_{pp}}{2\pi p_\perp d p_\perp dy}
 \right) 
\end{eqnarray}
is as small as possible.
 The 
parameters  of the fit are related to  $N_{core}(c)$ and $N_{corona}(c)$ 
defining   the two-component interaction region
\begin{eqnarray}
\alpha(c)&=&\frac{N_{core}({0\mbox{-}5\%})}{N_{core}(c)} \nonumber \\
\beta(c)&=&\frac{N_{core}({0\mbox{-}5\%})}{2 N_{core}(c)}N_{corona}(c)
-\frac{N_{corona}({0\mbox{-}5\%})}{2} \ .
\end{eqnarray}
The parameters $\alpha(c)$ and $\beta(c)$ are taken the same for 4 kinds of
 identified particles analyzed $K^+$, $K^-$, $p$ and $\bar{p}$. 
We 
did not include pions in the fit. Pions are less sensitive to the
 transverse flow and their thermal and $p\mbox{-}p$ spectra are similar. 
Therefore the
fit is not sensitive to the separation between the core and the
 corona in that case.
In Figs. \ref{fig:kaons} and \ref{fig:pap} are shown the spectra transformed 
according to Eq. (\ref{eq:albe}). The procedure of making all spectra look
 similar after subtraction of the  $p\mbox{-}p$ 
contribution works moderately well
 for kaons and very well for protons and antiprotons. The thermal 
component of 
the particle emission can be described using one thermal distribution 
for all centralities. For comparison in the lower 
part in  Figs.  \ref{fig:kaons} and \ref{fig:pap} we show the 
experimental data for different centralities scaled by the number
 of participating nucleon pairs. It proves that spectra measured for 
different centralities have different slopes, and after integration
 would not give the same number of produced kaons, protons or antiprotons 
per participating pair. Only after subtracting the corona
 contribution to the spectra a universal thermal emission is recovered.

\begin{figure}
\includegraphics[width=.48\textwidth]{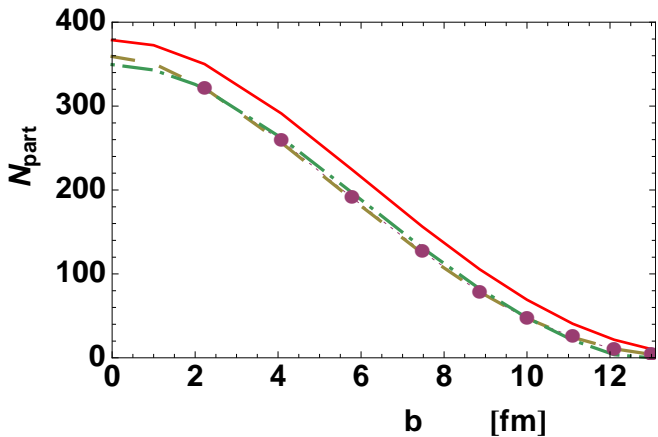}
\caption{(Color online) Dependence of the source size 
 on the impact parameter for the Glauber optical model (Eq. \ref{eq:wn}) 
(solid line),
 for the thermal fireball defined by the higher density in transverse plane 
 (Eq. \ref{eq:wndens})
(dashed-dotted line),  and for the thermal fireball defined by 
nucleons undergoing more than one collision (dashed line). The dots
are the results of the fit for the size of the thermal core at different
 impact parameters.}
\label{fig:sourcesize}
\end{figure}

\begin{figure}
\includegraphics[width=.4\textwidth]{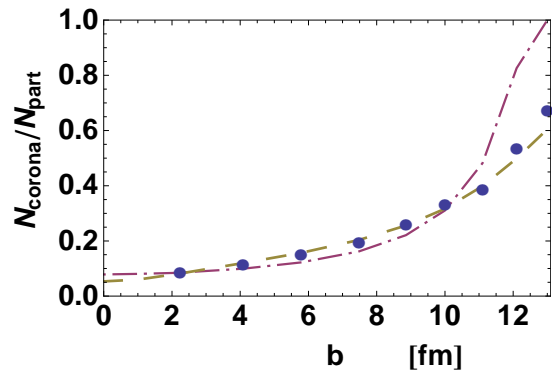}
\caption{(Color online) Ratio of the corona size to all participants 
 for the thermal fireball defined by the higher density in transverse plane 
 (Eq. \ref{eq:wndens})
(dashed-dotted line),  and for the thermal fireball defined by 
nucleons undergoing more than one collision (dashed line). The dots
are the results of the fit for the size of the thermal core at different
 impact parameters.}
\label{fig:sourceprop}
\end{figure}

From the fit, parameters $N_{core}$ can be extracted for each centrality.
The procedure needs an input for $N_{corona}(0\mbox{-}5\%)$, we take the value 
$N_{corona}(0\mbox{-}5\%)=29$ suggested by the core-corona models and by
 the analysis of \cite{Bozek:2005eu}. The fitted  source sizes
 as function of the impact parameter are shown by the dots
 in Fig. \ref{fig:sourcesize}. The thermal fireball size $N_{core}(c)$ is 
clearly smaller than the whole interaction region given by $N_{part}$ from 
a Glauber model calculation (solid line in Fig.\ref{fig:sourcesize}). 

 In the Figure are also shown prediction of two models of the core 
(details of the calculation are presented in the next section.). 
The first one (dashed-dotted line) 
is defined by the minimal density of participant nucleons in the
transverse plane $\rho(x,y)>\rho_{crit}$ 
\begin{equation}
N_{core}=\int dx dy \rho(x,y) \Theta(\rho(x,y)-\rho_{crit}) \ .
\end{equation}
The cut-off density is $\rho_{crit}=1.2$fm$^{-2}$. It is smaller than the value
 $\rho_{crit}=2$fm$^{-2}$ used in Ref.  \cite{Bozek:2005eu}, 
but in that previous analysis part of the particles emitted
 from the corona have been assumed to be 
 absorbed by the core. The dashed line in Fig. \ref{fig:sourcesize} 
represents the core size defined as nucleons that underwent more than 
one collisions. Counting only the nucleons with $N_{coll}>1$ as belonging to 
the core can explain the strangeness suppression in peripheral collisions
\cite{Becattini:2008yn}. In Fig. \ref{fig:sourceprop} is shown the 
proportion of the core size to the size of the whole interaction region.
Both the model based on the minimal density 
(Glauber model with high density) and the model where 
 the core size is defined by nucleons with multiple 
collisions (Glauber model with $N_{coll}>1$) work well
for central and semiperipheral collisions $b<10$fm.
The last model gives correct estimates of the core 
size also for more peripheral collisions.
 The two-component analysis
 of the spectra of identified particles produced at different centralities
shows that the emission occurs in two sources within the interaction
 region. The separation between the two components is somewhat arbitrary. 
Qualitatively with increasing impact parameter the corona 
portion of the interaction region increases and the spectra 
change from thermal like more to N-N collisions like
 (Fig. \ref{fig:sourceprop}). We quote two models 
of the core and corona that give reasonable estimate of their sizes.

\section{Dense thermal source}

 \label{sec:dense}

The initial distribution of matter in the transverse  plane (${\bf s}=(x,y)$)
in heavy-ion collisions can be described using the optical 
Glauber model \cite{Bialas:1976ed}.
Nucleons in the two colliding nuclei are distributed according to the
 Woods-Saxon density
\begin{equation}
\label{eq:woodssaxon}
\rho(r)=\frac{\rho_0}{1+\exp\left((r-R_A)/a\right)} .
\end{equation}
For the Au nucleus ($A=197$)
 we take \cite{Abelev:2007qg}, the central density
$\rho_0=0.161$fm$^{-3}$, the radius $R_A=6.5$fm and the parameter $a=0.535$fm.
The nuclear thickness function is defined as
\begin{equation}
T_A({\bf s})=\int_{-\infty}^\infty \rho(\sqrt{{\bf s}^2+z^2}) dz .
\end{equation}
For the collision of a symmetric system at impact parameter $b$,
 the density of participants in the 
transverse plane is
\begin{eqnarray}
& & \rho_w({\bf s})=T_A({\bf s}+b/2)\left(1-(1-
\frac{\sigma T_A({\bf s}-b/2)}{A})^A\right)\nonumber \\  & &
 + T_A({\bf s}-b/2)\left(1-(1-
\frac{\sigma T_A({\bf s}+b/2)}{A})^A\right) \ ,
\end{eqnarray}
where $\sigma=41$mb is the N-N cross section.
In the optical Glauber model the number of participants is given by
\begin{equation}
\label{eq:wn}
N_{part}=\int {\bf d^2 s} \rho_w({\bf s}) \ 
\end{equation}
and can be calculated for each impact parameter (solid line in Fig.
 \ref{fig:sourcesize}).

\begin{figure}
\includegraphics[width=.35\textwidth]{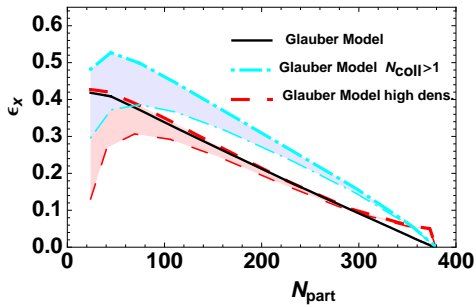}
\caption{(Color online) 
Spatial eccentricity of the initial state from the Glauber model
 (solid line), from the Glauber model with the density of participants
 $>1.2$fm$^{-2}$ (upper dashed line), same but including a 
weighted contribution of isotropic emission from the corona
 (lower dashed line), 
from the Glauber model  including only participants with more 
than one collision (upper dashed-dotted line),  same but including 
a weighted contribution of the isotropic emission from the corona 
(lower dashed-dotted line).}
\label{fig:exini}
\end{figure}

In the first model \cite{Bozek:2005eu} of the dense core, the thermal fireball 
is restricted to densities above a given minimal density $\rho_{crit}$
\begin{equation}
\label{eq:wndens}
\rho_w^{dens}({\bf s})=\rho_w({\bf s})\Theta(\rho_w({\bf s})-\rho_{crit})
\end{equation}
and the core size as measured by the number of participants is
\begin{equation}
\label{eq:ndens}
N_{core}^{dens}=\int {\bf d^2 s} \rho_w^{dens}({\bf s}) \ .
\end{equation}
$N_w^{dens}$ as function of the impact parameter is shown by the dashed-dotted 
line in Fig. \ref{fig:sourcesize} for  $\rho_{crit}=1.2$fm$^{-2}$.

The second model \cite{Becattini:2008yn}
 of the corona is based on the condition that nucleons
 undergo more than one collision. The transverse plane density of participants  
with more than one collision is obtained after a simple calculation
\begin{eqnarray}
\label{eq:wncoll}
& & \rho_w^{coll}({\bf s})=T_A({\bf s}+b/2)\left(1-(1-
\frac{\sigma T_A({\bf s}-b/2)}{A})^{A-1}\right.\nonumber \\ 
& & \left.(1+\sigma T_A({\bf s}-b/2)\frac{A-1}{A})\right) \nonumber \\  & &
 + T_A({\bf s}-b/2)\left(1-(1-
\frac{\sigma T_A({\bf s}+b/2)}{A})^{A-1}\right.\nonumber \\ 
& & \left.(1+\sigma T_A({\bf s}+b/2)\frac{A-1}{A})\right) \ .
\end{eqnarray}
The size of the core $N_{core}^{coll}$  (dashed-dotted line in Fig. 
\ref{fig:sourcesize}) is defined analogously to (\ref{eq:ndens}).

At nonzero impact parameter the interaction region is azimuthally asymmetric.
The spatial eccentricity is defined as
\begin{equation}
\epsilon_x=\frac{\int {\bf d^2 s} (y^2-x^2)\rho_w({\bf s})}{\int {\bf d^2 s}
 (x^2+y^2)\rho_w({\bf s})} .
\end{equation}
The above formula can be used for all three densities of the source defined 
above $\rho_w$, $\rho_w^{dens}$ and $\rho_w^{coll}$. The results are
 shown in Fig. \ref{fig:exini}.
This definition of the  eccentricity is the so called standard
 eccentricity \cite{Miller:2003kd},  that 
gives
zero for head-on collisions. During the expansion, the initial eccentricity
is transformed into the momentum eccentricity of the collective flow 
\cite{Voloshin:2006gz}
 and eventually into the azimuthal asymmetry of the transverse momentum 
emission of particles.  In the linear approximation, 
the final elliptic flow coefficient is proportional to  $\epsilon_x$ 
\cite{Broniowski:2007ft}. This argument should be amended because of the 
emission from the corona. In the first approximation we assume that the
 emission from the corona is isotropic. 
In the experiment, particles from the core
 (anisotropic emission) and from the corona are summed, this reduces the
 total anisotropy.
On the average the number of particles
 emitted per participant pair from the core 
is a factor $\alpha=1.65$ larger  than from 
the corona  \cite{Bozek:2005eu}.
The weighted eccentricity taking into account the contribution from the corona 
can be defined as
\begin{equation}
\label{eq:wex}
\frac{\alpha N_{core}}{\alpha N_{core}+N_{corona}}\epsilon_x \ .
\end{equation}
 The reduced, weighted 
eccentricities are show by the lower dashed and dashed-dotted curves in Fig. 
\ref{fig:exini}. The bands between the lower and upper curves 
in Fig. \ref{fig:exini} denote the spread between the core eccentricity and the 
corresponding weighted eccentricity (\ref{eq:wex}). Since some 
anisotropy in the emission from the corona is possible due to
 a shadowing effect of the core, we consider the spread in the
band as the uncertainty of the prediction of the effective eccentricities.

\begin{figure}
\includegraphics[width=.3\textwidth]{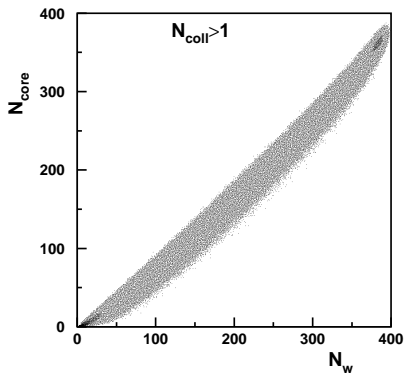}
\caption{Histogram of  event by event realizations of the
 number of participants in the core for a fixed number of all participants.}
\label{fig:dnhist}
\end{figure}

\begin{figure}
\includegraphics[width=.35\textwidth]{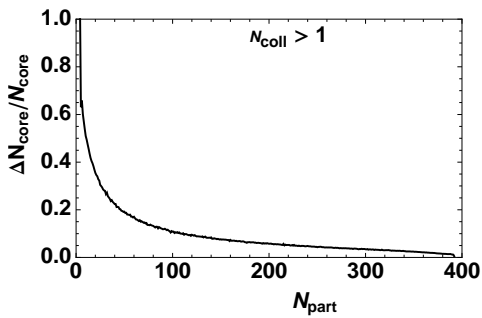}
\caption{Scaled standard deviation of the size of the core as function 
of the number of all participants.}
\label{fig:dn}
\end{figure}

A Monte-Carlo realization of the Glauber model can be 
 used \cite{Broniowski:2007nz}.
In the two nuclei the positions of  $A$ nucleons are generated according to the 
Woods-Saxon distribution (\ref{eq:woodssaxon}). In the generation of nulceon 
positions a minimal distance of $0.4$fm between two nucleons is imposed.
Two nuclei composed of $A$ nucleons are shifted by $\pm b/2$ in the $x$ 
direction
 in the transverse plane and straight line trajectories of all the  nucleons are
 followed.  A N-N collisions 
occurs if two nucleons from different nuclei pass within  the 
distance $\sqrt{\sigma/\pi}$ from each other in the transverse plane. The
 number of nucleons colliding at least once (all participant 
 nucleons) is recorded, the same for the number of nucleons that 
collided more than once (nucleons in the core). Their positions in the 
transverse plane are summed to give the densities $\rho_w({\bf s})$ 
and $\rho_{core}^{coll}({\bf s})$.  Event by event correlation between 
$N_{part}$ and $N_{core}$ is shown in Fig. \ref{fig:dnhist}.  
For more 
peripheral collisions (and for Cu-Cu 
collisions) event by event 
fluctuation of the core size become important (Fig. \ref{fig:dn}) 
and could modify the 
conclusions on the collective flow and final spectra. For smaller impact
 parameters the relative importance of the fluctuations of the core is smaller
and an average  of the core size (density) over the events can be used as 
an initial state of the thermal fireball.

\begin{figure}
\includegraphics[width=.35\textwidth]{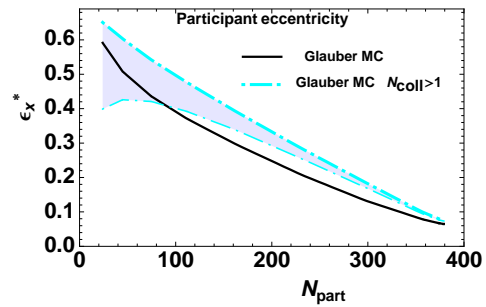}
\caption{(color online) Participant eccentricity in the initial state
 from the Glauber model including 
event by event eccentricity fluctuations in the distribution of 
all participants (solid line),  
of participants with more than one collision(upper dashed-dotted curve), same
  but including 
a weighted contribution of the isotropic emission from the corona 
(lower dashed-dotted line).}
\label{fig:exrini}
\end{figure}

\begin{figure}
\includegraphics[width=.35\textwidth]{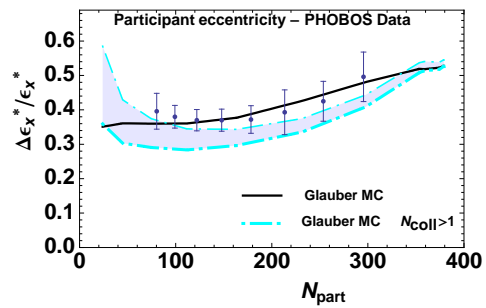}
\caption{(Color online) 
Scaled standard deviation of the participant eccentricity for all
 participants (solid line), for
 participants with more than one collision(lower dashed-dotted curve), same
  but including 
a weighted contribution of the isotropic emission from the corona 
(upper dashed-dotted line). }
\label{fig:dexrini}
\end{figure}

Event by event fluctuations are important for
 the spatial eccentricity of the initial distribution of collision 
centers in the transverse plane \cite{Aguiar:2001ac,Miller:2003kd}.
In each event the coordinates in the transverse plane are rotated 
to maximize the eccentricity \cite{Broniowski:2007nz}. 
Even at zero-impact parameter fluctuations 
of the distribution in the finite number of collision points give 
non-zero eccentricity. We calculate the eccentricity in this way 
(the so called participant eccentricity) for the Glauber Monte-Carlo model and
for the core defined as nucleons with more than one collision. The eccentricity
is larger than in the standard definition (Fig. \ref{fig:exrini}). 
At zero impact parameter the participant eccentricity 
is nonzero (and similar) both for all 
participant nucleons and for nucleons from the core. 
For intermediate impact parameters, 
participant eccentricity is larger for participants in the core 
 even after the reduction 
of the effect by the emission from the corona (Eq. \ref{eq:wex}).

To relate the initial spatial eccentricity to the elliptic flow coefficient 
$v_2$ of the observed hadrons a dynamical calculation must be performed. 
A quantity that is more directly related to the observed particle spectra 
is the scaled standard deviation of the elliptic flow. The scaled 
standard deviation in the initial state $\Delta \epsilon^\star/\epsilon^\star$
can be compared to measured scaled fluctuations of the elliptic flow 
$\Delta v_2/v_2$
 \cite{Alver:2008zz}.
Glauber Monte-Carlo model results are very close
 to the measured scaled elliptic
flow coefficient \cite{Alver:2008zz,Voloshin:2006gz,Broniowski:2007ft} 
(solid line in Fig. \ref{fig:dexrini}).
For the core-corona scenario the denominator 
of the  ratio $\Delta \epsilon^\star/\epsilon^\star$  is increased,
in the numerator we neglect the small contribution from fluctuations of the 
isotropic corona emission. Thus, the scaled standard deviation is reduced for 
the core-corona scenario. This reduction is smaller if the weighting
 (\ref{eq:wex})
 is applied to the anisotropy of the core.
 In a more detailed analysis presented in the 
next section we show that the interplay of the core and corona emission 
depends on the particle type and the transverse momentum. Since
 thermal spectra are harder than  spectra from  N-N emission 
in the corona, the
 elliptic flow at intermediate $p_\perp$ is dominated by the core contribution.

\section{Hydrodynamic expansion of the thermal source}
\label{sec:hydro}

\begin{figure}
\includegraphics[width=.5\textwidth]{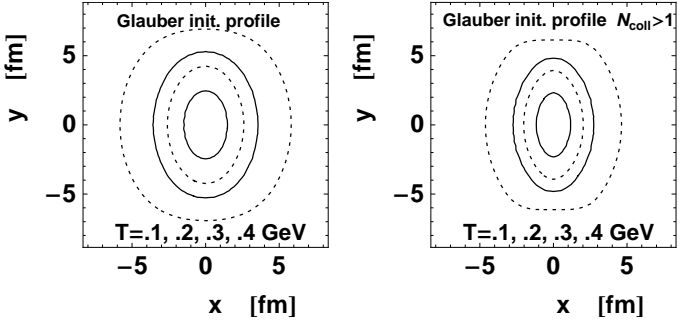}
\caption{Contour plot of the initial temperature profile ($b=10$fm) 
when the 
 entropy density  is proportional 
to the participant density from the  Glauber model
 (left panel) or  to the density of participants with more than one collision 
(right panel).}
\label{fig:iniprofile}
\end{figure}

Matter in the  dense fireball created in the first stage of the collision 
flows in the longitudinal direction. We assume a boost invariant Bjorken scaling
expansion 
 in the beam direction. In the transverse direction there 
is no flow initially, but later, pressure gradients lead to transverse
 flow and expansion. The expansion can be modelled as ideal
 fluid hydrodynamics in the dense thermalized phase 
\cite{Kolb:2003dz,Huovinen:2006jp,Nonaka:2007nn}, followed by 
particle emission at the freeze-out. We solve $2\mbox{+}1$-dimensional 
relativistic hydrodynamic equations
\begin{equation}
\partial_\mu T^{\mu \nu}=0 \ ,
\end{equation}
with the energy-momentum tensor of the form
 $T^{\mu \nu}=(\epsilon+p)u^\mu u^\nu-g^{\mu \nu}p$.
With boost-invariance in the longitudinal direction the four velocity is 
\begin{equation}
u^\mu=(\frac{t}{\sqrt{t^2-z^2}}\gamma_\perp,v_x\gamma_\perp,
v_y\gamma_\perp,\frac{z}{\sqrt{t^2-z^2}}\gamma_\perp) , 
\end{equation}
with 
$\gamma_\perp=1/\sqrt{1-v_x^2-v_y^2}$. The energy density $\epsilon$, the 
 pressure $p$ and the 
velocities $v_x$, $v_y$ depend on the  transverse coordinates and on
 the proper time $\sqrt{t^2-z^2}$. Moreover, the energy density and the 
pressure are related by the EOS.
We take a realistic parameterization thereof
  \cite{Chojnacki:2007jc}. The use of such a realistic EOS 
has been shown to be indispensable for a quantitative description of RHIC 
measurements
\cite{Chojnacki:2007rq,Broniowski:2008vp}. Interpolated lattice data
 above the critical temperature of 
$T_c = 170$MeV and an EOS of  noninteracting hadrons
 at lower temperatures are taken. The limiting formulas are joined without a 
soft point.

\begin{figure}
\includegraphics[width=.3\textwidth]{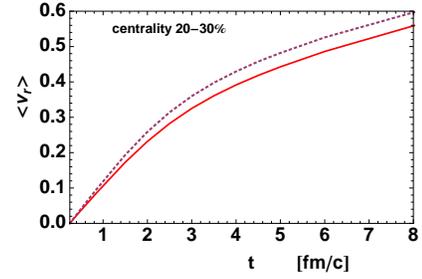}
\caption{(Color online) 
Average transverse velocity (\ref{eq:vsr}) as function of the time of
 the 
hydrodynamic expansion for standard Glauber initial profile (solid line) 
and for  the initial profile including only nucleons with more than one 
collision (dotted line).}
\label{fig:vsr}
\end{figure}

We solve numerically hydrodynamic equations for two different 
initial conditions, 
the Glauber Model initial density (Eq. \ref{eq:wndens})
and the thermal core including nucleons with more than one collision 
(Eq. \ref{eq:wncoll}). The entropy density in the fireball created at impact 
parameter $b$ is assumed to be proportional 
to density of participants
\begin{equation}
s({\bf s},b)= s_0 \frac{\rho_w({\bf s},b)}{\rho_w({\bf 0},0)}
\end{equation}
The entropy density $s_0$ at the center of the fireball in central 
collisions corresponds to a temperature of $500$MeV.
For the core scenario we take $\rho_w^{coll}$ to define
 the profile of the entropy and the constant $s_0$
 is rescaled to get the same total entropy (for the corona we assume an 
effective entropy a factor $1/\alpha$ smaller per participant). 
Two set of calculations are performed at impact parameters $b=2.23, \ 4.08, \
5.78, \ 7.48, \ 8.86, \ 10.0, \ 11.1, \ 12.1$fm corresponding to centrality bins
from $0-5$\% to $60-70$\%
\cite{Abalev:2008ez}.
An early initial time of the hydrodynamic evolution is chosen 
$\tau_0=0.25$fm/c \cite{Chojnacki:2007rq}.

\begin{figure}
\includegraphics[width=.3\textwidth]{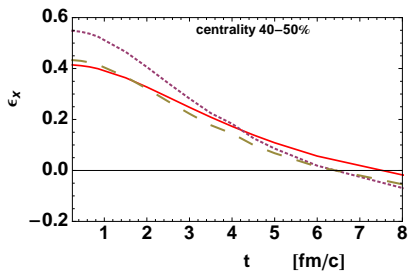}
\caption{(color online) 
Time evolution of the spatial anisotropy (\ref{eq:exhy}) for the 
standard Glauber initial profile (solid line), for  the initial profile 
including only nucleons with more than one 
collision (dotted line), and the same but 
including a weighted contribution from the 
isotropic emission from the corona (dashed line).}
\label{fig:exhy}
\end{figure}

In Fig. \ref{fig:iniprofile} is presented a contour plot
 of the initial temperature for $b=10$fm.
 The profile for the dense core is steeper than for
the standard Glauber model. Also the eccentricity is larger for
 nucleons undergoing multiple collisions.
Larger gradients of the pressure make the acceleration faster, 
 the average transverse velocity 
\begin{equation}
\label{eq:vsr}
<v_{r}>=\frac{\int dx dy \gamma_\perp
 v_\perp \epsilon}{\int dx dy \gamma_\perp \epsilon}
\end{equation}
increases faster in the expansion
 of the core (Fig. \ref{fig:vsr}). 
Spatial eccentricity 
\begin{equation}
\label{eq:exhy}
\epsilon_x=\frac{\int dx dy  (y^2-x^2)\gamma_\perp \epsilon}{\int dx dy  
(x^2+y^2)\gamma_\perp \epsilon}
\end{equation}
decreases
during the evolution. The initial eccentricity of the dense core is larger than
 for  the standard Glauber initial density, but faster expansion of the former 
makes  the difference disappear (Fig. \ref{fig:exhy}).
 If the weighting factor (\ref{eq:wex}) is included 
the effective spatial  eccentricity is similar in the two scenarios.

\begin{figure}
\includegraphics[width=.4\textwidth]{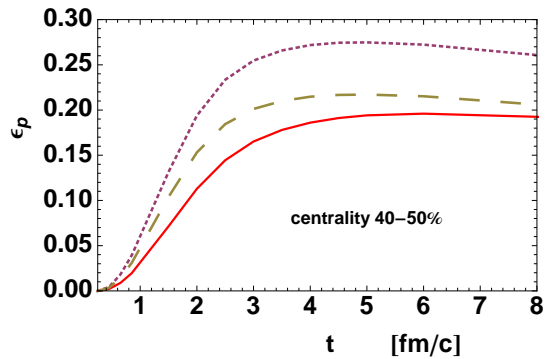}
\caption{(Color online) 
Time evolution of the momentum anisotropy (\ref{eq:ephy})
 for the  standard Glauber initial profile (solid line), 
for  the initial profile 
including only nucleons with more than one 
collision (dotted line), and the same but 
including a weighted contribution from the 
isotropic emission from the corona (dashed line).}
\label{fig:ephy}
\end{figure}

More important for the final particle spectra is the momentum anisotropy
\begin{equation}
\label{eq:ephy}
\epsilon_p=\frac{\int dx dy (T_{xx}-T_{yy})}{\int dx dy  (T_{xx}+T_{yy})} \ .
\end{equation}
Larger gradients in the in-plane direction cause a stronger 
collective flow in that direction. 
Spatial eccentricity disappears, but imprints the velocity field of the fluid
 (Fig. \ref{fig:ephy}). The momentum anisotropy is stronger
 if the initial state is the thermal core, the effect survives the reduction 
by the  weighting factor (\ref{eq:wex}).

\begin{figure}
\includegraphics[width=.3\textwidth]{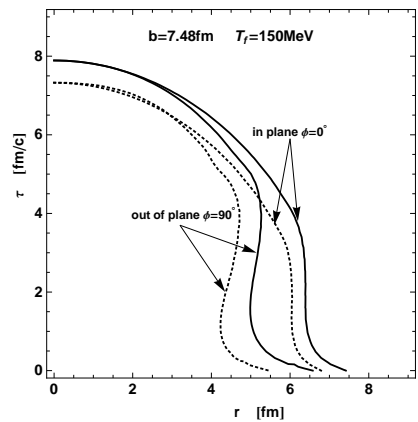}
\caption{The freeze-out hypersurface $T(t,r,\phi)=150$MeV 
in $t\mbox{-}r$ coordinates  for in plane $\phi=0^{\deg}$ 
and out of plane  $\phi=90^{\deg}$ directions, for the standard Glauber
 initial condition (solid lines) and for the initial condition including 
only nucleons 
with more than one 
collision (dashed lines) at impact parameter $b=7.48$fm. }
\label{fig:freeze}
\end{figure}

The hydrodynamic evolution is stopped at the freeze-out temperature 
$T_f=150$MeV. The smaller size and larger temperature
 gradients in the core make it 
freeze out earlier than the standard 
fireball composed of all the participant nucleons.
Earlier freeze-out means a shorter expansion and less
 time for the build up of the flow, this partially  reduces the effect of 
faster acceleration in the expansion from the core.
From the freeze-out hypersurface $\Sigma$ particles are emitted according 
to the Cooper-Frye formula \cite{Cooper:1974mv}
\begin{equation}
E\frac{dN}{d^3p}=\int_\Sigma d\Sigma_\mu p^\mu f(p^\mu u_\mu) \ ,
\end{equation}
where the integration is over the hypersurface elements 
$d\Sigma_\mu$ and $f(p_\mu u^\mu)$ is the thermal (Bose or Fermi) distribution.
The procedure of particle emission from the hypersurface and of the decay of 
 resonances is realized using the statistical emission code THERMINATOR  
\cite{Kisiel:2005hn}. Final $\pi$, $K$ and $p$ spectra
and the  elliptic flow coefficients are calculated.

\begin{figure}
\includegraphics[width=.4\textwidth]{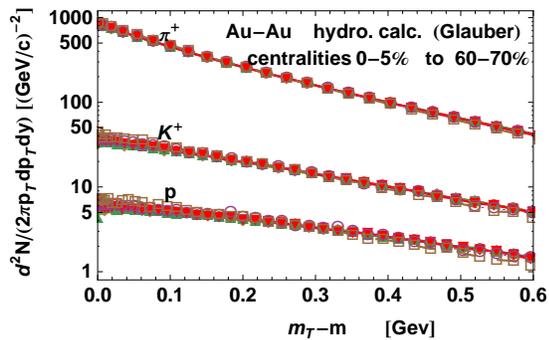}
\caption{(Color online) 
Transverse mass spectra of $\pi$, $K$ and $p$ from 
the hydrodynamic evolution of standard Glauber initial conditions 
with a freeze-out at $T_f=150$MeV. 
Results for  8
 centrality classes
 ($0\mbox{-}5\%$, $5\mbox{-}10\%$, \dots , $60\mbox{-}70\%$) scaled by the
mean number of participant pairs are shown.}
\label{fig:hyspe}
\end{figure}

\begin{figure}
\includegraphics[width=.4\textwidth]{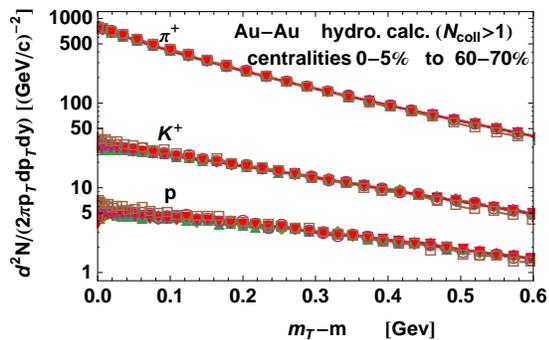}
\caption{(Color online) 
 Same as Fig. \ref{fig:hyspe} but for the dense core initial condition
($N_{coll}>1$).}
\label{fig:hyspecbec}
\end{figure}

\begin{figure}
\includegraphics[width=.3\textwidth]{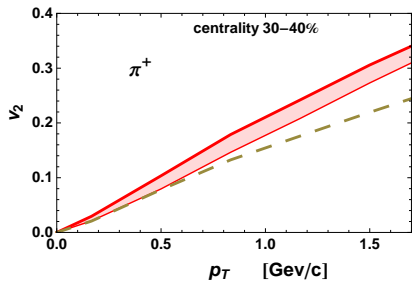}
\caption{(Color online) 
Elliptic flow coefficient $v_2$ as function of transverse momentum
 for pions, for the  standard Glauber initial condition (dashed line), for the
 dense core ($N_{coll}>1$) initial condition (upper solid line) 
and the same but including the weighted isotropic 
contribution from the emission from the corona (Eq. \ref{eq:redpt})
 (lower solid line).}
\label{fig:v2pi}
\end{figure}


\begin{figure}
\includegraphics[width=.3\textwidth]{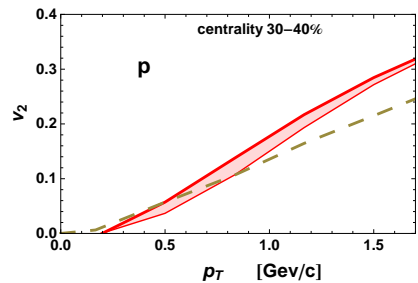}
\caption{(Color online) Same as Fig. \ref{fig:v2pi} but for protons.}
\label{fig:v2pr}
\end{figure}

In Figs. \ref{fig:hyspe} and \ref{fig:hyspecbec} are shown particle
 spectra for the two initial  densities considered. The 
spectra are scaled by the number of all participants or
 by the number of participants in the core 
respectively.
If the initial state  is  the thermal core,  spectra are slightly
 harder than for the standard Glauber initial conditions. Comparing 
spectra obtained at different centralities, we notice that the amount of 
transverse flow is a bit larger for central collisions. This violates
 to some extend the assumption made in the fit separating the observed 
spectra into the  thermal and  direct components (Sect. \ref{sec:fit}).
The separation between the core and corona components  presented 
is  approximate since the thermal component  depends weakly on
 the centrality of the collision.
The elliptic flow coefficient as function of $p_\perp$ for pions 
and protons in shown in Figs. \ref{fig:v2pi} and \ref{fig:v2pr}.
The elliptic flow for particles emitted after the expansion of the
 core is larger than the elliptic flow calculated for standard Glauber 
initial conditions.
In Sect. \ref{sec:dense} an average reduction factor for the elliptic flow 
taking into account the emission from the corona is introduced.
For the momentum dependent coefficient $v_2$ we use the following reduction 
factor
\begin{equation}
\label{eq:redpt}
\frac{N_{core} \frac{dN_{th}}{2\pi p_\perp dp_\perp } }
{N_{core} \frac{dN_{th}}{2\pi p_\perp dp_\perp }+ N_{corona} 
\frac{dN_{pp}}{2\pi p_\perp dp_\perp }}v_2(p_\perp) \ .
\end{equation}
The strongest reduction occurs for small momenta. At larger $p_\perp$ the
 harder thermal spectra dominate. Thus the difference between 
the upper solid lines (without the reduction factor (\ref{eq:redpt}))
 and the lower solid lines (with the reduction factor) in Figs. \ref{fig:v2pi} 
and \ref{fig:v2pr} is small at intermediate $p_\perp$. The average 
reduction factor (\ref{eq:wex}) overestimates the effect, especially
 for heavier particles. The difference between hydrodynamic expansions from 
the two  initial densities 
is most visible in the final differential elliptic flow in $p_\perp$. 
We expect
this effect to be even stronger for smaller colliding systems.

\section{Conclusions}

We discuss the scenario of a two component source in the
 interaction region in ultrarelativistic nuclear collisions.
The interaction region includes a 
 core  composed of dense, thermalized matter, that evolves 
collectively and of an outer corona where particles are  emitted in sparse
 N-N collisions  directly. This idea has been proposed earlier
 in order to explain the increase of particle multiplicity and strangeness 
content with centrality \cite{Bozek:2005eu}. In this first estimate 
the core has been  defined as the part of the interaction region with
 a sufficiently large 
density of participants. A different definition of the core is 
proposed in Ref. \cite{Becattini:2008yn}. Only nucleons that collided more
 than once emit particles that thermalize. The rest of the participants 
constitutes the corona. We fit the spectra of kaons, protons and antiprotons 
at different centralities with  two components, a thermal part
 and a contribution from N-N collisions. The extracted size of 
the thermal core is compared to the prediction of the two models
 of the core-corona separation. The 
size of the core can be defined by the nucleons in the high density region 
for impact parameters $b<10$fm,
whereas 
the number of 
 nucleons that underwent several collisions is a good
 estimate of the size of the thermal core for all centralities.

We give a 
formula describing the density of nucleons that collided more than once. 
We calculate 
 the spatial eccentricity of the models of the dense 
dense core. For the core composed of nucleons with more than one 
collisions it is 
larger than for the standard Glauber model fireball. 
We expect 
a strong elliptic flow after the expansion of such a dense, asymmetric core.
From a Glauber Monte-Carlo model calculation we find that 
scaled eccentricity fluctuations are slightly reduced for the 
emission from the core compared to estimates taking 
into account all participants \cite{Miller:2003kd}.

The hydrodynamic evolution starting with the standard Glauber 
model density of the fireball is compared to  the expansion of the
core composed of nucleons with multiple collisions. The density of the
 core has larger gradients and larger 
spatial eccentricity. It leads to a faster
 expansion, stronger transverse and elliptic flows. Even after taking 
into account the emission from the corona, the final elliptic flow for 
the core-corona fireball is larger than the one for the standard Glauber 
model fireball. This result shows that the study 
the scaling of the elliptic flow with 
the size of the system should involve also the possibility of a 
two component emission with proportions changing with centrality.

\section*{Acknowledgments}
The author is grateful to Miko\l aj Chojnacki for discussions  on the
hydrodynamic calculations and the freeze-out procedure.
\bibliography{../hydr}

\begin{thebibliography}{36}
\expandafter\ifx\csname natexlab\endcsname\relax\def\natexlab#1{#1}\fi
\expandafter\ifx\csname bibnamefont\endcsname\relax
  \def\bibnamefont#1{#1}\fi
\expandafter\ifx\csname bibfnamefont\endcsname\relax
  \def\bibfnamefont#1{#1}\fi
\expandafter\ifx\csname citenamefont\endcsname\relax
  \def\citenamefont#1{#1}\fi
\expandafter\ifx\csname url\endcsname\relax
  \def\url#1{\texttt{#1}}\fi
\expandafter\ifx\csname urlprefix\endcsname\relax\def\urlprefix{URL }\fi
\providecommand{\bibinfo}[2]{#2}
\providecommand{\eprint}[2][]{\url{#2}}

\bibitem[{\citenamefont{Adams et~al.}(2005)}]{Adams:2005dq}
\bibinfo{author}{\bibfnamefont{J.}~\bibnamefont{Adams}} \bibnamefont{et~al.}
  (\bibinfo{collaboration}{STAR}), \bibinfo{journal}{Nucl. Phys.}
  \textbf{\bibinfo{volume}{A757}}, \bibinfo{pages}{102} (\bibinfo{year}{2005}),
  \eprint{nucl-ex/0501009}.

\bibitem[{\citenamefont{Adcox et~al.}(2005)}]{Adcox:2004mh}
\bibinfo{author}{\bibfnamefont{K.}~\bibnamefont{Adcox}} \bibnamefont{et~al.}
  (\bibinfo{collaboration}{PHENIX}), \bibinfo{journal}{Nucl. Phys.}
  \textbf{\bibinfo{volume}{A757}}, \bibinfo{pages}{184} (\bibinfo{year}{2005}),
  \eprint{nucl-ex/0410003}.

\bibitem[{\citenamefont{Arsene et~al.}(2005)}]{Arsene:2004fa}
\bibinfo{author}{\bibfnamefont{I.}~\bibnamefont{Arsene}} \bibnamefont{et~al.}
  (\bibinfo{collaboration}{BRAHMS}), \bibinfo{journal}{Nucl. Phys.}
  \textbf{\bibinfo{volume}{A757}}, \bibinfo{pages}{1} (\bibinfo{year}{2005}),
  \eprint{nucl-ex/0410020}.

\bibitem[{\citenamefont{Back et~al.}(2005)}]{Back:2004je}
\bibinfo{author}{\bibfnamefont{B.~B.} \bibnamefont{Back}} \bibnamefont{et~al.}
  (\bibinfo{collaboration}{PHOBOS}), \bibinfo{journal}{Nucl. Phys.}
  \textbf{\bibinfo{volume}{A757}}, \bibinfo{pages}{28} (\bibinfo{year}{2005}),
  \eprint{nucl-ex/0410022}.

\bibitem[{\citenamefont{Schnedermann et~al.}(1993)\citenamefont{Schnedermann,
  Sollfrank, and Heinz}}]{Schnedermann:1993ws}
\bibinfo{author}{\bibfnamefont{E.}~\bibnamefont{Schnedermann}},
  \bibinfo{author}{\bibfnamefont{J.}~\bibnamefont{Sollfrank}},
  \bibnamefont{and} \bibinfo{author}{\bibfnamefont{U.~W.} \bibnamefont{Heinz}},
  \bibinfo{journal}{Phys. Rev.} \textbf{\bibinfo{volume}{C48}},
  \bibinfo{pages}{2462} (\bibinfo{year}{1993}), \eprint{nucl-th/9307020}.

\bibitem[{\citenamefont{Adams et~al.}(2004)}]{Adams:2003xp}
\bibinfo{author}{\bibfnamefont{J.}~\bibnamefont{Adams}} \bibnamefont{et~al.}
  (\bibinfo{collaboration}{STAR}), \bibinfo{journal}{Phys. Rev. Lett.}
  \textbf{\bibinfo{volume}{92}}, \bibinfo{pages}{112301}
  (\bibinfo{year}{2004}), \eprint{nucl-ex/0310004}.

\bibitem[{\citenamefont{Andronic et~al.}(2006)\citenamefont{Andronic,
  Braun-Munzinger, and Stachel}}]{Andronic:2005yp}
\bibinfo{author}{\bibfnamefont{A.}~\bibnamefont{Andronic}},
  \bibinfo{author}{\bibfnamefont{P.}~\bibnamefont{Braun-Munzinger}},
  \bibnamefont{and} \bibinfo{author}{\bibfnamefont{J.}~\bibnamefont{Stachel}},
  \bibinfo{journal}{Nucl. Phys.} \textbf{\bibinfo{volume}{A772}},
  \bibinfo{pages}{167} (\bibinfo{year}{2006}), \eprint{nucl-th/0511071}.

\bibitem[{\citenamefont{Becattini et~al.}(2006)\citenamefont{Becattini,
  Manninen, and Gazdzicki}}]{Becattini:2005xt}
\bibinfo{author}{\bibfnamefont{F.}~\bibnamefont{Becattini}},
  \bibinfo{author}{\bibfnamefont{J.}~\bibnamefont{Manninen}}, \bibnamefont{and}
  \bibinfo{author}{\bibfnamefont{M.}~\bibnamefont{Gazdzicki}},
  \bibinfo{journal}{Phys. Rev.} \textbf{\bibinfo{volume}{C73}},
  \bibinfo{pages}{044905} (\bibinfo{year}{2006}), \eprint{hep-ph/0511092}.

\bibitem[{\citenamefont{Cleymans et~al.}(2005)\citenamefont{Cleymans, Kampfer,
  Kaneta, Wheaton, and Xu}}]{Cleymans:2004pp}
\bibinfo{author}{\bibfnamefont{J.}~\bibnamefont{Cleymans}},
  \bibinfo{author}{\bibfnamefont{B.}~\bibnamefont{Kampfer}},
  \bibinfo{author}{\bibfnamefont{M.}~\bibnamefont{Kaneta}},
  \bibinfo{author}{\bibfnamefont{S.}~\bibnamefont{Wheaton}}, \bibnamefont{and}
  \bibinfo{author}{\bibfnamefont{N.}~\bibnamefont{Xu}}, \bibinfo{journal}{Phys.
  Rev.} \textbf{\bibinfo{volume}{C71}}, \bibinfo{pages}{054901}
  (\bibinfo{year}{2005}), \eprint{hep-ph/0409071}.

\bibitem[{\citenamefont{Florkowski et~al.}(2002)\citenamefont{Florkowski,
  Broniowski, and Michalec}}]{Florkowski:2001fp}
\bibinfo{author}{\bibfnamefont{W.}~\bibnamefont{Florkowski}},
  \bibinfo{author}{\bibfnamefont{W.}~\bibnamefont{Broniowski}},
  \bibnamefont{and} \bibinfo{author}{\bibfnamefont{M.}~\bibnamefont{Michalec}},
  \bibinfo{journal}{Acta Phys. Polon.} \textbf{\bibinfo{volume}{B33}},
  \bibinfo{pages}{761} (\bibinfo{year}{2002}), \eprint{nucl-th/0106009}.

\bibitem[{\citenamefont{Rafelski et~al.}(2005)\citenamefont{Rafelski,
  Letessier, and Torrieri}}]{Rafelski:2004dp}
\bibinfo{author}{\bibfnamefont{J.}~\bibnamefont{Rafelski}},
  \bibinfo{author}{\bibfnamefont{J.}~\bibnamefont{Letessier}},
  \bibnamefont{and} \bibinfo{author}{\bibfnamefont{G.}~\bibnamefont{Torrieri}},
  \bibinfo{journal}{Phys. Rev.} \textbf{\bibinfo{volume}{C72}},
  \bibinfo{pages}{024905} (\bibinfo{year}{2005}), \eprint{nucl-th/0412072}.

\bibitem[{\citenamefont{Hirano et~al.}(2008)\citenamefont{Hirano, van~der Kolk,
  and Bilandzic}}]{Hirano:2008hy}
\bibinfo{author}{\bibfnamefont{T.}~\bibnamefont{Hirano}},
  \bibinfo{author}{\bibfnamefont{N.}~\bibnamefont{van~der Kolk}},
  \bibnamefont{and} \bibinfo{author}{\bibfnamefont{A.}~\bibnamefont{Bilandzic}}
  (\bibinfo{year}{2008}), \eprint{arXiv:0808.2684 [nucl-th]}.

\bibitem[{\citenamefont{Huovinen and Ruuskanen}(2006)}]{Huovinen:2006jp}
\bibinfo{author}{\bibfnamefont{P.}~\bibnamefont{Huovinen}} \bibnamefont{and}
  \bibinfo{author}{\bibfnamefont{P.~V.} \bibnamefont{Ruuskanen}},
  \bibinfo{journal}{Ann. Rev. Nucl. Part. Sci.} \textbf{\bibinfo{volume}{56}},
  \bibinfo{pages}{163} (\bibinfo{year}{2006}), \eprint{nucl-th/0605008}.

\bibitem[{\citenamefont{Kolb and Heinz}(2004)}]{Kolb:2003dz}
\bibinfo{author}{\bibfnamefont{P.~F.} \bibnamefont{Kolb}} \bibnamefont{and}
  \bibinfo{author}{\bibfnamefont{U.~W.} \bibnamefont{Heinz}}, in
  \emph{\bibinfo{booktitle}{Quark Gluon Plasma 3}}, edited by
  \bibinfo{editor}{\bibfnamefont{R.}~\bibnamefont{Hwa}} \bibnamefont{and}
  \bibinfo{editor}{\bibfnamefont{X.~N.} \bibnamefont{Wang}}
  (\bibinfo{publisher}{World Scientific, Singapore}, \bibinfo{year}{2004}),
  \eprint{nucl-th/0305084}.

\bibitem[{\citenamefont{Nonaka}(2007)}]{Nonaka:2007nn}
\bibinfo{author}{\bibfnamefont{C.}~\bibnamefont{Nonaka}}, \bibinfo{journal}{J.
  Phys.} \textbf{\bibinfo{volume}{G34}}, \bibinfo{pages}{S313}
  (\bibinfo{year}{2007}), \eprint{nucl-th/0702082}.

\bibitem[{\citenamefont{Broniowski et~al.}(2008)\citenamefont{Broniowski,
  Chojnacki, Florkowski, and Kisiel}}]{Broniowski:2008vp}
\bibinfo{author}{\bibfnamefont{W.}~\bibnamefont{Broniowski}},
  \bibinfo{author}{\bibfnamefont{M.}~\bibnamefont{Chojnacki}},
  \bibinfo{author}{\bibfnamefont{W.}~\bibnamefont{Florkowski}},
  \bibnamefont{and} \bibinfo{author}{\bibfnamefont{A.}~\bibnamefont{Kisiel}},
  \bibinfo{journal}{Phys. Rev. Lett.} \textbf{\bibinfo{volume}{101}},
  \bibinfo{pages}{022301} (\bibinfo{year}{2008}), \eprint{arXiv:0801.4361
  [nucl-th]}.

\bibitem[{\citenamefont{Chojnacki et~al.}(2008)\citenamefont{Chojnacki,
  Florkowski, Broniowski, and Kisiel}}]{Chojnacki:2007rq}
\bibinfo{author}{\bibfnamefont{M.}~\bibnamefont{Chojnacki}},
  \bibinfo{author}{\bibfnamefont{W.}~\bibnamefont{Florkowski}},
  \bibinfo{author}{\bibfnamefont{W.}~\bibnamefont{Broniowski}},
  \bibnamefont{and} \bibinfo{author}{\bibfnamefont{A.}~\bibnamefont{Kisiel}},
  \bibinfo{journal}{Phys. Rev.} \textbf{\bibinfo{volume}{C78}},
  \bibinfo{pages}{014905} (\bibinfo{year}{2008}), \eprint{arXiv:0712.0947
  [nucl-th]}.

\bibitem[{\citenamefont{Kisiel et~al.}(2008)\citenamefont{Kisiel, Broniowski,
  Chojnacki, and Florkowski}}]{Kisiel:2008ws}
\bibinfo{author}{\bibfnamefont{A.}~\bibnamefont{Kisiel}},
  \bibinfo{author}{\bibfnamefont{W.}~\bibnamefont{Broniowski}},
  \bibinfo{author}{\bibfnamefont{M.}~\bibnamefont{Chojnacki}},
  \bibnamefont{and}
  \bibinfo{author}{\bibfnamefont{W.}~\bibnamefont{Florkowski}}
  (\bibinfo{year}{2008}), \eprint{arXiv:0808.3363 [nucl-th]}.

\bibitem[{\citenamefont{Voloshin}(2007)}]{Voloshin:2007af}
\bibinfo{author}{\bibfnamefont{S.~A.} \bibnamefont{Voloshin}}
  (\bibinfo{collaboration}{STAR}), \bibinfo{journal}{J. Phys.}
  \textbf{\bibinfo{volume}{G34}}, \bibinfo{pages}{S883} (\bibinfo{year}{2007}),
  \eprint{nucl-ex/0701038}.

\bibitem[{\citenamefont{Chajecki and Lisa}(2008)}]{Chajecki:2008yi}
\bibinfo{author}{\bibfnamefont{Z.}~\bibnamefont{Chajecki}} \bibnamefont{and}
  \bibinfo{author}{\bibfnamefont{M.}~\bibnamefont{Lisa}}
  (\bibinfo{year}{2008}), \eprint{arXiv:0807.3569 [nucl-th]}.

\bibitem[{\citenamefont{Bozek}(2005)}]{Bozek:2005eu}
\bibinfo{author}{\bibfnamefont{P.}~\bibnamefont{Bozek}}, \bibinfo{journal}{Acta
  Phys. Polon.} \textbf{\bibinfo{volume}{B36}}, \bibinfo{pages}{3071}
  (\bibinfo{year}{2005}), \eprint{nucl-th/0506037}.

\bibitem[{\citenamefont{Pantuev}(2007)}]{Pantuev:2005jt}
\bibinfo{author}{\bibfnamefont{V.~S.} \bibnamefont{Pantuev}},
  \bibinfo{journal}{JETP Lett.} \textbf{\bibinfo{volume}{85}},
  \bibinfo{pages}{104} (\bibinfo{year}{2007}), \eprint{hep-ph/0506095}.

\bibitem[{\citenamefont{Werner}(2007)}]{Werner:2007bf}
\bibinfo{author}{\bibfnamefont{K.}~\bibnamefont{Werner}},
  \bibinfo{journal}{Phys. Rev. Lett.} \textbf{\bibinfo{volume}{98}},
  \bibinfo{pages}{152301} (\bibinfo{year}{2007}), \eprint{0704.1270}.

\bibitem[{\citenamefont{Becattini and Manninen}(2008)}]{Becattini:2008yn}
\bibinfo{author}{\bibfnamefont{F.}~\bibnamefont{Becattini}} \bibnamefont{and}
  \bibinfo{author}{\bibfnamefont{J.}~\bibnamefont{Manninen}},
  \bibinfo{journal}{J. Phys.} \textbf{\bibinfo{volume}{G35}},
  \bibinfo{pages}{104013} (\bibinfo{year}{2008}), \eprint{arXiv:0805.0098
  [nucl-th]}.

\bibitem[{\citenamefont{Abalev et~al.}(2008)}]{Abalev:2008ez}
\bibinfo{author}{\bibfnamefont{B.}~\bibnamefont{Abalev}} \bibnamefont{et~al.}
  (\bibinfo{collaboration}{STAR}) (\bibinfo{year}{2008}),
  \eprint{arXiv:0808.2041 [nucl-ex]}.

\bibitem[{\citenamefont{Bia\l{}as et~al.}(1976)\citenamefont{Bia\l{}as,
  Bleszy\'nski, and Czy\.z}}]{Bialas:1976ed}
\bibinfo{author}{\bibfnamefont{A.}~\bibnamefont{Bia\l{}as}},
  \bibinfo{author}{\bibfnamefont{M.}~\bibnamefont{Bleszy\'nski}},
  \bibnamefont{and} \bibinfo{author}{\bibfnamefont{W.}~\bibnamefont{Czy\.z}},
  \bibinfo{journal}{Nucl. Phys.} \textbf{\bibinfo{volume}{B111}},
  \bibinfo{pages}{461} (\bibinfo{year}{1976}).

\bibitem[{\citenamefont{Abelev et~al.}(2007)}]{Abelev:2007qg}
\bibinfo{author}{\bibfnamefont{B.~I.} \bibnamefont{Abelev}}
  \bibnamefont{et~al.} (\bibinfo{collaboration}{STAR}), \bibinfo{journal}{Phys.
  Rev.} \textbf{\bibinfo{volume}{C75}}, \bibinfo{pages}{054906}
  (\bibinfo{year}{2007}), \eprint{nucl-ex/0701010}.

\bibitem[{\citenamefont{Miller and Snellings}(2003)}]{Miller:2003kd}
\bibinfo{author}{\bibfnamefont{M.}~\bibnamefont{Miller}} \bibnamefont{and}
  \bibinfo{author}{\bibfnamefont{R.}~\bibnamefont{Snellings}}
  (\bibinfo{year}{2003}), \eprint{nucl-ex/0312008}.

\bibitem[{\citenamefont{Voloshin}(2006)}]{Voloshin:2006gz}
\bibinfo{author}{\bibfnamefont{S.~A.} \bibnamefont{Voloshin}}
  (\bibinfo{year}{2006}), \eprint{nucl-th/0606022}.

\bibitem[{\citenamefont{Broniowski
  et~al.}(2007{\natexlab{a}})\citenamefont{Broniowski, Bozek, and
  Rybczynski}}]{Broniowski:2007ft}
\bibinfo{author}{\bibfnamefont{W.}~\bibnamefont{Broniowski}},
  \bibinfo{author}{\bibfnamefont{P.}~\bibnamefont{Bozek}}, \bibnamefont{and}
  \bibinfo{author}{\bibfnamefont{M.}~\bibnamefont{Rybczynski}},
  \bibinfo{journal}{Phys. Rev.} \textbf{\bibinfo{volume}{C76}},
  \bibinfo{pages}{054905} (\bibinfo{year}{2007}{\natexlab{a}}),
  \eprint{arXiv:0706.4266 [nucl-th]}.

\bibitem[{\citenamefont{Broniowski
  et~al.}(2007{\natexlab{b}})\citenamefont{Broniowski, Rybczynski, and
  Bozek}}]{Broniowski:2007nz}
\bibinfo{author}{\bibfnamefont{W.}~\bibnamefont{Broniowski}},
  \bibinfo{author}{\bibfnamefont{M.}~\bibnamefont{Rybczynski}},
  \bibnamefont{and} \bibinfo{author}{\bibfnamefont{P.}~\bibnamefont{Bozek}}
  (\bibinfo{year}{2007}{\natexlab{b}}), \eprint{arXiv:0710.5731 [nucl-th]}.

\bibitem[{\citenamefont{Aguiar et~al.}(2002)\citenamefont{Aguiar, Hama, Kodama,
  and Osada}}]{Aguiar:2001ac}
\bibinfo{author}{\bibfnamefont{C.~E.} \bibnamefont{Aguiar}},
  \bibinfo{author}{\bibfnamefont{Y.}~\bibnamefont{Hama}},
  \bibinfo{author}{\bibfnamefont{T.}~\bibnamefont{Kodama}}, \bibnamefont{and}
  \bibinfo{author}{\bibfnamefont{T.}~\bibnamefont{Osada}},
  \bibinfo{journal}{Nucl. Phys.} \textbf{\bibinfo{volume}{A698}},
  \bibinfo{pages}{639} (\bibinfo{year}{2002}), \eprint{hep-ph/0106266}.

\bibitem[{\citenamefont{Alver et~al.}(2008)}]{Alver:2008zz}
\bibinfo{author}{\bibfnamefont{B.}~\bibnamefont{Alver}} \bibnamefont{et~al.},
  \bibinfo{journal}{Phys. Rev.} \textbf{\bibinfo{volume}{C77}},
  \bibinfo{pages}{014906} (\bibinfo{year}{2008}), \eprint{0711.3724}.

\bibitem[{\citenamefont{Chojnacki and Florkowski}(2007)}]{Chojnacki:2007jc}
\bibinfo{author}{\bibfnamefont{M.}~\bibnamefont{Chojnacki}} \bibnamefont{and}
  \bibinfo{author}{\bibfnamefont{W.}~\bibnamefont{Florkowski}},
  \bibinfo{journal}{Acta Phys. Polon.} \textbf{\bibinfo{volume}{B38}},
  \bibinfo{pages}{3249} (\bibinfo{year}{2007}), \eprint{nucl-th/0702030}.

\bibitem[{\citenamefont{Cooper and Frye}(1974)}]{Cooper:1974mv}
\bibinfo{author}{\bibfnamefont{F.}~\bibnamefont{Cooper}} \bibnamefont{and}
  \bibinfo{author}{\bibfnamefont{G.}~\bibnamefont{Frye}},
  \bibinfo{journal}{Phys. Rev.} \textbf{\bibinfo{volume}{D10}},
  \bibinfo{pages}{186} (\bibinfo{year}{1974}).

\bibitem[{\citenamefont{Kisiel et~al.}(2006)\citenamefont{Kisiel, Taluc,
  Broniowski, and Florkowski}}]{Kisiel:2005hn}
\bibinfo{author}{\bibfnamefont{A.}~\bibnamefont{Kisiel}},
  \bibinfo{author}{\bibfnamefont{T.}~\bibnamefont{Taluc}},
  \bibinfo{author}{\bibfnamefont{W.}~\bibnamefont{Broniowski}},
  \bibnamefont{and}
  \bibinfo{author}{\bibfnamefont{W.}~\bibnamefont{Florkowski}},
  \bibinfo{journal}{Comput. Phys. Commun.} \textbf{\bibinfo{volume}{174}},
  \bibinfo{pages}{669} (\bibinfo{year}{2006}), \eprint{nucl-th/0504047}.

\end{thebibliography}

\end{document}